**Association of night-waking and inattention/hyperactivity symptoms trajectories in preschool-aged children**

**Authors:**
Eve Reynaud PhD,[1,2,3] Anne Forhan,[1,2] Barbara Heude PhD,[1,2] Marie-Aline Charles MD,[1,2] and Sabine Plancoulaine MD, PhD,[1,2]

**Affiliations:**
[1] INSERM, UMR1153, Epidemiology and Statistics Sorbonne Paris Cité Research Center (CRESS), early ORigins of Child Health And Development Team (ORCHAD), Villejuif, F-94807 France;
[2] Paris Descartes University, Sorbonne Paris Cité, Paris, France.
[3] Ecole des Hautes Etudes en Santé Publique (EHESP), Rennes, F-35043, France

**Address correspondence to:**
Sabine Plancoulaine, INSERM U1153, Team 6 ORCHAD, 16 Avenue Paul Vaillant Couturier, 94807 Villejuif Cedex, France, [sabine.plancoulaine@inserm.fr], + 33 145-595-109

**Funding :**
No funding was received

**Competing interests statement:**
No competing interests to be declared

**Data availability statement:**
The data that support the findings of this study are available from the corresponding author SP upon reasonable request but restrictions apply to the availability of these data, which were used under license for the current study, and so are not publicly available.

**ABSTRACT**

**Objective:** To study the longitudinal associations between inattention/hyperactivity symptoms and night-waking in preschool-years, in light of their joint evolution.

**Study design:** Within the French birth-cohort study EDEN, repeated measures of 1342 children's night-waking and inattention/hyperactivity symptoms were collected at age 2, 3 and 5-6 through questionnaires. Trajectories were computed using group-based modeling. Logistic regressions, adjusted for confounding factors, were used to measure the association between trajectories and to determine risk factors for belonging to the identified joint trajectories.

**Results:** Two night-waking trajectories were observed, 20% of the children had a trajectory of "common night-waking", and 80% a trajectory of "rare night-waking". The children were distributed in three inattention/hyperactivity trajectories, a low (47%), medium (40%) and high one (13%). Both night-waking and inattention/hyperactivity trajectories showed persistence of difficulties in preschool years. The risk of presenting a high inattention/hyperactivity trajectory compared to a low one was of 4.19[2.68-6.53] for common night-wakers, compared to rare night-wakers. Factors associated with joint trajectories were parent's education level and history of childhood behavioral problems, and the child's gender, night-sleep duration and collective care at 2 years of age.

**Conclusion:** Results suggest that children presenting behavioral difficulties would benefit from a systematic investigation of their sleep quality and conversely.

# INTRODUCTION

Night-waking is very common amongst preschool-aged children, with 36% signaling their awakening at least once a night.[1] It is one of the primary complaints of parents to pediatricians[2] and has been shown to be persistent within preschool years.[3] Early inattention/hyperactivity symptoms have also been shown to be persistent over this age range[4] and predictive of later poor cognitive development[5] and academic underachievement.[6,7] However, little is known about the associations between night-waking and inattention/hyperactivity in preschool years. In light of the early onset and persistence of night waking and inattention/hyperactivity symptoms, it is essential to study both behaviors in early development.

A systematic literature review[8] found 6 studies investigating the association between night-waking and inattention/hyperactivity symptoms in preschool years[9–14] all with a cross sectional design. Half reported that more night-waking was associated with higher inattention/hyperactivity symptoms.[10,12,14] In the Hatzinger *et al* study,[11] this association was observed in boys only, while Hall *et al*[9] and Lehmkulh *et al*[13] found no significant association. In a previous longitudinal study conducted in the EDEN birth-cohort[15] including 1342 children, we showed that children presenting a common night-waking trajectory from age 2 to 5-6 had higher scores of attention/hyperactivity symptoms at age 5-6, compared to children presenting a rare night-waking trajectory. As associations between sleep and behavior may be bidirectional,[16] we hypothesized that the associations between night-waking and inattention/hyperactivity symptoms could be better understood by studying their joint evolution throughout preschool years.
Thus, the objective of the present study was to investigate the joint trajectories of night-waking and inattention/hyperactivity symptoms from age 2 to 5-6, and their associations with parental and child characteristics, using group-based trajectory modeling.

**METHODS**

**Study population**

The EDEN birth cohort study recruited pregnant women before their 24th week of amenorrhea between 2003 and 2006 in two French university hospitals, Poitiers and Nancy[17]. Inclusion criteria were for women to be older than 18 years old, to have no history of diabetes, to be functionally literate in French and able to give informed consent. Additional inclusion criteria were to be under social security coverage and to have no plan of changing address within the next 3 years. Women with multiple pregnancies were excluded.

Among the 3758 women invited to participate, 2002 (53%) agreed to enroll in the study. Due to miscarriages, stillbirths and attrition, 1899 children were included at birth. Written informed consent was obtained twice from parents: at enrolment for themselves and after the child's birth for their child. The study was approved by the research ethics committee of Bicêtre Hospital (Comité Consultatif de Protection des Personnes dans la Recherche Biomédicale) and by the Data Protection Authority (Commission Nationale de l'Informatique et des Libertés). All methods were performed in accordance with relevant guidelines and regulations. Further details on the EDEN study protocol have been published elsewhere.[17]

**Measures**

*Inattention/hyperactivity symptoms*

The child's inattention/hyperactivity score was obtained at three time points, when the child was 2, 3 and 5-6 years old. At age 2, parents completed a paper-based questionnaire including 6 questions, where they stated whether their child never, sometimes, or frequently displayed the following behaviors over the last 3 months: 1) couldn't stay in place, was agitated or hyperactive,

2) was distracted, had difficulties completing an activity, 3) was constantly squirming, 4) couldn't concentrate or maintain his/her attention over a long time span, 5) had difficulties waiting for his turn during a game, 6) was inattentive. A score between 0 and 12 was obtained, with 12 the highest inattention/hyperactivity score. At age 3 and 5-6, parents completed the Strengths and Difficulties Questionnaire (SDQ).[18] It is a commonly used questionnaire[11,12,14] that has been validated in preschool-aged children[19] and in a French non-clinical pediatric population.[20] Parents had to state whether those 5 statements concerning their child were not true, somewhat true or certainly true over the last 6 months: 1) restless, overactive, cannot stay still for long, 2) constantly fidgeting or squirming, 3) easily distracted, concentration wanders, 4) thinks things out before acting, 5) good attention span, sees work through to the end. A score between 0 and 10 was obtained, with 10 the highest inattention/hyperactivity score. To ensure comparability between inattention/ hyperactivity scores throughout the follow-up period, raw scores were transformed into z-scores, using the within population mean and standard deviation at each age.

*Night-waking*

In the same paper-based questionnaire, parents reported how many times a week their child woke up during the night at age 2, 3 and 5-6. In accordance with the literature,[21,22] "frequent night waking" was defined as the child waking every other night or more, as opposed to "no or occasional night waking".

*Covariates*

The child's characteristics included gender, birth order (first/other) and term at birth (weeks of amenorrhea). Parental characteristics were household income at inclusion (less than 1500€, between 1500 and 3000€ or more than 3000€ per month), education (defined as the highest number of years of study reached by one of both parents), history of parental behavior problems in childhood (one or both parents who declared to have had a behavioral problem during childhood), mother's age at delivery, mother's depressive symptoms (Center for epidemiologic studies depression scale (CES-D) score of 23 or more, threshold validated in French women[23]), smoking tobacco during pregnancy. The child's lifestyle and sleep at age 2 included the number of hours spent in front of the television per day, main care arrangement (in large collective settings like preschool or day care centers vs. childminder), the child's night-sleep duration at age 2, and the child's sleep habits which were defined as inadvisable if the child had an irregular sleep time or fell asleep with parental presence (yes/no).

**Statistical analyses**

Children were excluded from the analyses if information on both night-waking and inattention/hyperactivity was missing for more than one time point, resulting in a sample size of 1342 children (71% of the children included at birth). Amongst them, missing data for the covariates represented only 2.0% of the total data, thus simple imputation were performed (i.e. mean value for continuous variables and the modal value for categorical ones). To optimize the analysis of repeated measures in a context of co-evolving outcomes, Nagin's method for group-based trajectory modeling[24] (PROC TRAJ procedure, SAS 9.4 SAS institute INC, Cary, NC, USA ®) was used to compute joint trajectories of night-waking and inattention/hyperactivity from age 2 to 5-6. This method allowed us to identify within our population distinctive developmental trajectories for both outcomes. The model provides for each child his/her probability of membership to the different trajectories. It also provides the probability of each combination of night-waking and inattention/hyperactivity trajectories, called joint probabilities and the conditional probabilities, which are the probability of belonging to a night-waking trajectory knowing the inattention/hyperactivity trajectory, and vice-versa. The joint and conditional probabilities are weighted by the individual probability of membership to the different trajectories. We attributed for each child the trajectory of night-waking and the trajectory of inattention/hyperactivity for which he/she had the highest probability of belonging. We used logistic regressions to assess the associations between trajectories, unadjusted and adjusted for the recruitment center (Poitiers or Nancy) and the covariates described above. Using adjusted multinomial logistic regressions, we then investigated the risk factors in association with the joint trajectories.

**RESULTS**

**Population description**

Children included in the analyses, compared to those who were not, came from families with higher income (31% versus 19% with a total income > 3000€ per months, $p<0.01$) and higher education (1.3 years mean difference, $p<0.01$). Included mothers also showed less depressive symptoms during pregnancy (7% versus 14%, $p<0.01$). The population characteristics are described in Table 1. Mothers were on average 30 years old at delivery and 47% of the children were girls. At the age of 2, children watched on average 43 minutes of television per day. As previously reported[3], the optimal trajectory model for describing night-waking patterns was a two-group model (Figure 1, panel A). The first trajectory, showing low and decreasing prevalence of night-waking over the years was labelled "rare night-waking" represented 80% (N=1073) of the population. The second trajectory, labelled "common night-waking" trajectory, represented the remaining 20% (N=269) of the population and was characterized by a peak in night-waking prevalence at age 3. Three trajectories of inattention/hyperactivity were identified (Figure 1, panel B), labelled low, medium and high inattention/hyperactivity trajectories, representing respectively 47% (N=630), 40% (N=538) and 13% (N=174) of the population. The three trajectories were distinct from the beginning, and remained relatively steady throughout the follow-up, reflecting a perseverance of the inattention/hyperactivity score rank in preschool years. However, the gap between children belonging to the high inattention/hyperactivity trajectory and the others widened over the years.

**Associations between trajectories**

Table 2 describes the interrelationship between night-waking and inattention/hyperactivity trajectories from age 2 to 5-6 years, and provides the conditional and joint probabilities. The probability of having a high inattention/hyperactivity trajectory when belonging to the common night-waking trajectory was of 0.20, versus 0.13 ($p$=0.01) when belonging to the rare one. The probability of having a common night-waking trajectory when belonging to the high inattention/hyperactivity trajectory was of 0.31, versus 0.14 ($p$<0.01) when belonging to the low one.

Figure 2 provides the distributions of the trajectories after the child attribution to the trajectories for which he/she had the highest probability of belonging. In the panel A are reported the crude distribution of inattention/hyperactivity trajectories by night-waking trajectories and vice-versa for panel B. The distribution of inattention/hyperactivity trajectories significantly differed according to the night-waking trajectories ($p$<0.01). Amongst children belonging to the rare night-waking trajectory (panel A), more than half belonged to the low inattention/hyperactivity trajectory and 11% to the high inattention/hyperactivity one. In the common night waking trajectory, only 22% of the children belonged to the low inattention/hyperactivity trajectory and more than 20% to the high inattention/hyperactivity one. In panel B, the prevalence of children belonging to the common night-waking trajectory was similar within both the high and medium inattention/hyperactivity trajectories with respectively 32.2% and 28.6% (*p*=0.37). These prevalences were both significantly higher than the one observed in the low inattention/hyperactivity trajectory (*p*<0.01).

**Associations between joint trajectories and covariates**

The adjusted associations between the joint trajectories of night-waking and inattention/hyperactivity and covariates are reported in Table 3. The distributions of covariates by joint trajectories are in the supplementary data. Joint trajectories showed global associations with parental education and parent's history of behavior problems in childhood, child gender, collective care arrangement and night-sleep duration. Children who were first born, had shorter night sleep and had parents with a shorter education and more parental history of childhood behavioral problems were more likely to present a joint "common night-waking trajectory and high inattention/hyperactivity", compared to children in the joint "rare night-waking and low inattention/hyperactivity trajectories" (Reference group). Similarly, risk factors for belonging to the joint "rare night-waking high inattention/hyperactivity, compared to the reference group were to be a boy, have shorter night-sleep, to have for main care arrangement a collective care, to have a mother who smoked tobacco during pregnancy, parents with a shorter education and more parental history of childhood behavioral problems.

**Adjusted associations between trajectories**

Multivariate logistic regressions indicated that the associations between night-waking and inattention/hyperactivity trajectories described previously remained significant after adjustment for included covariates (Table 4). When comparing children with a medium or a high inattention/hyperactivity trajectory to those with a low one, the risk of belonging to the common night-waking trajectory were respectively of OR=3.73 CI 95% [2.66-5.23] (*p*<0.01) and OR=4.24 CI 95% [2.72-6.63] (*p*<0.01). Children belonging to the high inattention/hyperactivity trajectory compared to the medium one had equivalent risk of belonging to the common night-waking trajectory (*p*=0.51).

**DISCUSSION**

**Associations between trajectories**

The present study showed that both night-waking and inattention/hyperactivity symptoms were persistent through preschool years. Their trajectories between 2 and 5-6 were highly associated, even after adjusting for multiple confounding factors. These results are in accordance to those reported by Touchette *et al* between night sleep duration and hyperactivity.[16] In their study including 2057 children of the "Quebec longitudinal study of child development", they found that hyperactivity scores were persistent from age 1.5 to 5, and, except for 5% of the children, night sleep durations were also persistent. Nocturnal sleep durations and hyperactivity trajectories were also significantly associated ($X^2$=75.1, *p*<0.01).

**Associations between joint trajectories and covariates**

In the adjusted analyses, factors that were globally associated with joint trajectories were parental education, history of childhood behavioral problems, child gender, collective care and night-sleep duration. These results are in accordance with previous literature as male gender[25] and shorter parental education[26] are common predictors of hyperactivity symptoms. Associations have also been found with lower socio-economic status and night-waking[27].

*Socio-economic factors*

Shorter education often involves unfavorable prenatal conditions and environment for children as well as less access to education and medical care, which could explain the observed increase in behavioral difficulties for this group. The global association between income and joint trajectories was not significant. Yet, children growing in a wealthy family (>3000€/months), compared to those with a household income of less than 1500€/months, were less likely to belong to the joint "high inattention/hyperactivity and common night-waking" trajectory (the least favorable joint trajectory). The lack of global association could be due to a low variance in income in our study sample. The EDEN birth cohort population presents higher income then the general French population[17] and further selection due to missing data for night waking or inattention/hyperactivity deepened the miss representativeness of the sample.

*Parental history of behavioral problems*

History of childhood behavioral problems was associated independently of the socio-economic factors to joint trajectories. According to recent literature, the role of genetic in ADHD is substantial.[28] The genes that have been associated in the disease are mostly involved in the dopaminergic[29] and serotonergic[30] systems. Thus the independent association between the parent's childhood behavioral problems with the child joint trajectories, especially those involving medium or high inattention/hyperactivity trajectories, could be explained by genetics, under the hypothesis that genetics also plays a role in the lower spectrum of attention/hyperactivity difficulties. Another explanation for this association is that there might be unmeasured residual effects of the social environment.

*Gender*

Our results regarding gender were consistent with the literature, boys have repeatedly been shown to have greater risk of hyperactive symptoms in preschool years.[25] Touchette *et al* similarly found that risk factors of belonging to the least favorable joint trajectory (short-persistent nocturnal sleep duration and high hyperactivity) compared to the most favorable one (11h persistent nocturnal sleep duration and low hyperactivity) were to be a boy, and to have parents with a low education. But unlike our results, they also found that lower income was a risk factor of presenting the least favorable joint trajectory.

**Concordance with sleep duration**

As commented above, our findings with night-waking are very similar to those found in night-sleep duration, although our multivariate analyses were adjusted for this factor. Thus, sleep quality is associated with inattention/hyperactivity development independently of sleep quantity. Those results suggest that night-waking should be addressed in parallel to sleep quantity during consultations.

**Limitations**

The methodology used in our analysis allowed us to describe the joint evolution of night-waking and inattention/hyperactivity and their association, while taking into account confounding factors. Yet there are limitations to be noted in this study. The measurements used for night-waking and inattention/hyperactivity were both subjective. Thus the declared night awakenings were those noticed by parents, indicating the child's capacity to fall back asleep on his/her own. Also, parents who perceive their child's sleep as more problematic may also perceive their daytime behaviors as more problematic, thus the associations found are affected by the parent's belief of normative child behavior. We used a sub-scale of Strengths and Difficulties Questionnaire to assess inattention/hyperactivity. This test has shown good detection power for externalizing behavior when compared to a professionally conducted semi-structured interview[31] and has been validated in French children.[20] A second limit of this study is a relative lack of statistical power in the analysis of adjusted associations between covariates and the joint trajectories. Joint trajectories resulted in a six-group variable, with two including less than 60 children. Adjusting for confounding factors may help reducing bias but at the expense of robustness and statistical power. Thus we were not able to determine whether the lack of associations found with certain factors were due to a low statistical power.

**CONCLUSION**

There was a persistence of both night-waking and inattention/hyperactivity in preschool years, and the two behaviors co-evolve. Children who were first born, boys, from less educated families with a history of childhood behavior problems had a higher risk of presenting joint common night-waking and high inattention/hyperactivity trajectories. The results suggest that children presenting behavioral difficulties would benefit from a systematic investigation of their sleep quality and an adapted medical care such as cognitive behavioral therapy, and conversely.

**Acknowledgments**

Collaborators: We thank the EDEN mother-child cohort study group (I. Annesi-Maesano, J.Y. Bernard, J. Botton, M.A. Charles, P. Dargent-Molina, B. de Lauzon-Guillain, P. Ducimetière, M. de Agostini, B. Foliguet, A. Forhan, X. Fritel, A. Germa, V. Goua, R. Hankard, B. Heude, M. Kaminski, B. Larroquey, N. Lelong, J. Lepeule, G. Magnin, L. Marchand, C. Nabet, F. Pierre, R.

Slama, M.J. Saurel-Cubizolles, M. Schweitzer, O. Thiebaugeorges). We thank all funding sources for the EDEN study: Foundation for Medical Research (FRM, n° ARS-3.29), National Agency for Research (ANR, n° 03-BLAN-0359-01, n° 06-SEST-03501, n° 06-SEST-03502), National Institute For Research In Public Health (IRESP: TGIR cohorte santé 2008 program), French Ministry of Health (DGS, n° CV05000146), French Ministry of Research, INSERM Bone and Joint Diseases National Research (PRO-A) and Human Nutrition National Research Programs (n° 4NU06G), Paris-Sud University, Nestlé, French National Institute for Population Health Surveillance (InVS, n° 05-PCTT2043), French National Institute for Health Education (INPES, n° 007/05 DAS), the European Union FP7 programs, Diabetes National Research Program, Mutuelle Générale de l'Education Nationale complementary health insurance (MGEN), French national agency for food security, French speaking association for the study of diabetes and metabolism (ALFEDIAM).

**LIST OF ABBREVIATIONS**

CES-D   Center for epidemiologic studies depression scale

CI95%   Confidence Interval of 95 percent

SDQ    Strengths and Difficulties Questionnaire

OR    Odds Ratio

**FIGURES LEGENDS**

Figure 1. Trajectories of night-waking and inattention/hyperactivity amongst 1342 children aged 2 to 5-6 years, of the EDEN birth cohort. **a)** Frequent night-waking trajectories. White triangles represent the "common night-waking" trajectory (N=1076, 80%) and the white squares the "rare night-waking" one (N=269, 20%). **b)** Inattention/hyperactivity z-scores trajectories. The black triangles, circles and squares represent respectively the high (N=174, 13%), medium (N=538, 40%) and low (N=630, 47%) inattention/hyperactivity z-score trajectories.

Figure 2. Prevalences of inattention/hyperactivity and night-waking trajectories **a)** Distribution of inattention/hyperactivity trajectories by night-waking trajectories **b)** Distribution of night-waking trajectories by inattention/hyperactivity trajectories

**Author's contribution statement:**

Eve Reynaud carried out the analyses, interpreted the data, and drafted the initial manuscript. Anne Forhan participated in the data collection, managed the EDEN database and gave guidance for the statistical analyses. Barbara Heude PhD, Marie-Aline Charles MD MPH designed and coordinated the data collection and reviewed the manuscript. Sabine Plancoulaine MD PhD supervised the conception, design and interpretation of the analyses and critically reviewed the manuscript. All authors and approved the final manuscript as submitted.

**Table 1** Population characteristics (N=1342)

|  | % (N) | Mean (SD) |
|---|---|---|
| Parental characteristics | | |
|   Household income | | |
|     <1500 €/month | 11.3 (152) | |
|     1500-3000 €/month | 58.3 (782) | |
|     >3000 €/month | 30.4 (408) | |
|   Education (years)[a] | | 14.5 (2.4) |
|   History of childhood behavioral problem (yes) | 15.0 (201) | |
|   Maternal depression (CES-D ≥23)[b] | 6.8 (91) | |
|   Maternal age at delivery (years) | | 30.0 (4.7) |
|   Smoking during pregnancy (yes) | 20.9 (280) | |
| Child characteristics | | |
|   Perinatal factors | | |
|     Child gender (girl) | 47.5 (637) | |
|     First child (yes) | 46.7 (627) | |
|     Term at birth (weeks) | | 39.3 (1.7) |
|   Lifestyle and sleep at age 2 | | |
|     Collective care arrangement (yes) | 21.2 (285) | |
|     Television viewing (min/day) | | 42.6 (40.1) |
|     Inadvisable sleep habits (yes)[c] | 9.4 (126) | |
|     Night-sleep duration (h/day) | | 11.1 (0.77) |

[a] Number of years of schooling starting from first year of primary school e.g. 12 years corresponds to having completed high school
[b] Center for epidemiologic studies depression scale, cutoff validated in a French population, measured during pregnancy
[c] Irregular sleep time or falling asleep with parental presence

**Table 2** The interrelationship from age 2 to 5-6 between night-waking and inattention/hyperactivity trajectories (weighted by the individual probability of membership to each trajectory)

| A) Probability of inattention/hyperactivity trajectories conditional on night-waking trajectories | | | |
|---|---|---|---|
| | | Night-waking trajectory | |
| | | Rare | Common |
| Inattention/hyperactivity trajectory | Low | 0.51 [0.48-0.54] | 0.30 [0.25-0.36] |
| | Medium | 0.36 [0.33-0.39] | 0.49 [0.43-0.55] |
| | High | 0.13 [0.11-0.14] | 0.20 [0.16-0.25] |
| B) Probability of night-waking trajectories conditional on inattention/hyperactivity trajectories | | | |
| | | Inattention/hyperactivity trajectory | |
| | Low | Medium | High |
| Night-waking trajectory Rare | 0.86 [0.83-0.89] | 0.72 [0.69-0.76] | 0.69 [0.62-0.75] |
| Night-waking trajectory Common | 0.14 [0.11-0.17] | 0.28 [0.24-0.31] | 0.31 [0.25-0.38] |
| C) Joint probability of night-waking and inattention/hyperactivity | | | |
| | | Night-waking trajectory | |
| | | Rare | Common |
| Inattention/hyperactivity trajectory | Low | 0.40 [0.37-0.43] | 0.07 [0.05-0.08] |
| | Medium | 0.28 [0.26-0.31] | 0.11 [0.09-0.12] |
| | High | 0.10 [0.08-0.11] | 0.05 [0.03-0.06] |

**Table 3** Multivariate associations between joint trajectories and covariates (N=1342). The joint "rare night-waking and low inattention/hyperactivity trajectories (I/H)" is the reference

| | Rare night-waking (N=1073) | | | Common night-waking (N=269) | | | $p$[b] |
|---|---|---|---|---|---|---|---|
| | Low I/H[a] N=571 reference | Medium I/H N=384 OR [CI95%] | High I/H N=118 OR [CI95%] | Low I/H N=59 OR [CI95%] | Medium I/H N=154 OR [CI95%] | High I/H N=56 OR [CI95%] | |
| Parental characteristics | | | | | | | |
| Household income | | | | | | | 0.07 |
| <1500 €/month (ref) | | | | | | | |
| 1500-3000 €/month | | 0.94 [0.58-1.53] | 0.94 [0.49-1.83] | 1.19 [0.39-3.62] | 0.65 [0.36-1.18] | 0.51 [0.23-1.10] | |
| >3000 €/month | | 0.79 [0.45-1.37] | 0.78 [0.33-1.83] | 0.38 [0.10-1.36] | 0.64 [0.31-1.30] | **0.26 [0.08-0.85]** | |
| Education (years)[c] | | **0.91 [0.86-0.97]** | **0.80 [0.73-0.89]** | 1.10 [0.96-1.25] | **0.89 [0.82-0.97]** | **0.82 [0.71-0.94]** | **<0.01** |
| History of childhood behavioral problem (yes) | | **1.53 [1.04-2.25]** | **2.22 [1.31-3.77]** | 1.00 [0.43-2.33] | 1.62 [0.98-2.69] | **2.23 [1.11-4.49]** | **0.03** |
| Maternal depression (CES-D ≥23)[d] | | 1.15 [0.67-1.97] | 1.49 [0.70-3.18] | 0.93 [0.27-3.25] | 0.86 [0.40-1.84] | 1.36 [0.44-4.21] | 0.87 |
| Maternal age at delivery (year) | | 0.98 [0.95-1.01] | 0.96 [0.91-1.01] | 1.00 [0.93-1.07] | 0.99 [0.94-1.03] | 0.93 [0.87-1.00] | 0.28 |
| Smoking during pregnancy (yes) | | **1.43 [1.01-2.01]** | **1.74 [1.06-2.83]** | 1.15 [0.55-2.42] | **1.75 [1.12-2.73]** | 1.10 [0.55-2.21] | 0.09 |
| Child characteristics | | | | | | | |
| Perinatal factors | | | | | | | |
| Child gender (girl) | | **0.71 [0.54-0.93]** | **0.33 [0.21-0.52]** | 1.34 [0.77-2.33] | 0.74 [0.51-1.06] | 0.60 [0.33-1.08] | **<0.01** |
| First child (yes) | | 1.09 [0.81-1.46] | 1.19 [0.75-1.89] | 0.90 [0.49-1.65] | 0.89 [0.59-1.33] | **1.97 [1.02-3.82]** | 0.35 |
| Term at birth (week) | | 0.94 [0.86-1.01] | 1.05 [0.92-1.19] | 0.99 [0.83-1.19] | 0.91 [0.82-1.00] | 0.99 [0.84-1.17] | 0.23 |
| Lifestyle at age 2 | | | | | | | |
| Collective care (yes) | | 1.03 [0.74-1.41] | **0.39 [0.20-0.76]** | 1.21 [0.65-2.25] | 0.70 [0.43-1.14] | 0.52 [0.22-1.20] | **0.03** |
| Television viewing (h/day) | | **1.26 [1.01-1.56]** | 0.96 [0.69-1.33] | 1.42 [0.96-2.11] | 1.23 [0.93-1.62] | 1.24 [0.85-1.82] | 0.16 |
| Inadvisable sleep habits (yes) | | 1.17 [0.74-1.86] | 0.72 [0.33-1.58] | 0.70 [0.23-2.07] | 1.32 [0.73-2.37] | 0.40 [0.12-1.38] | 0.35 |
| Night-sleep duration (h/day) | | 0.89 [0.74-1.06] | **0.74 [0.56-0.98]** | **0.63 [0.43-0.91]** | **0.75 [0.58-0.95]** | **0.53 [0.36-0.77]** | **<0.01** |

Significant associations ($p<0.05$) are represented in bold
[a] I/H inattention/hyperactivity
[b] Global type 3 p-value, obtained from adjusted logistic regression
[c] Number of years of schooling starting from first year of primary school e.g. 12 years corresponds to having completed high school
[d] Center for epidemiologic studies depression scale. Cutoff validated in a French population. Measured during pregnancy

**Table 4** Adjusted associations between the inattention/hyperactivity trajectories and common night-waking trajectories

|  |  | OR [CI 95%] | p |
|---|---|---|---|
| Risk of presenting a medium inattention/hyperactivity compared to a low one |  |  |  |
|  | Rare night-waking | Reference |  |
|  | Common night-waking | 3.67 [2.62-5.14] | <0.01 |
| Risk of presenting a high inattention/hyperactivity compared to a low one |  |  |  |
|  | Rare night-waking | Reference |  |
|  | Common night-waking | 4.19 [2.68-6.53] | <0.01 |
| Risk of presenting a common night-waking trajectory compared to a rare one |  |  |  |
|  | Low inattention/hyperactivity trajectory | Reference |  |
|  | Medium inattention/hyperactivity trajectory | 3.73 [2.66-5.23] | <0.01 |
|  | High inattention/hyperactivity trajectory | 4.24 [2.72-6.63] | <0.01 |

Adjusted for recruitment center, household income, education, history of childhood behavioral problem, maternal depression, maternal age at delivery, smoking during pregnancy, child gender, birth order, term at birth, care arrangement, television viewing, sleep habits, night-sleep duration

Children belonging to the high inattention/hyperactivity trajectory compared to the medium one had equivalent risk of belonging to the common night-waking trajectory ($p$=0.51).

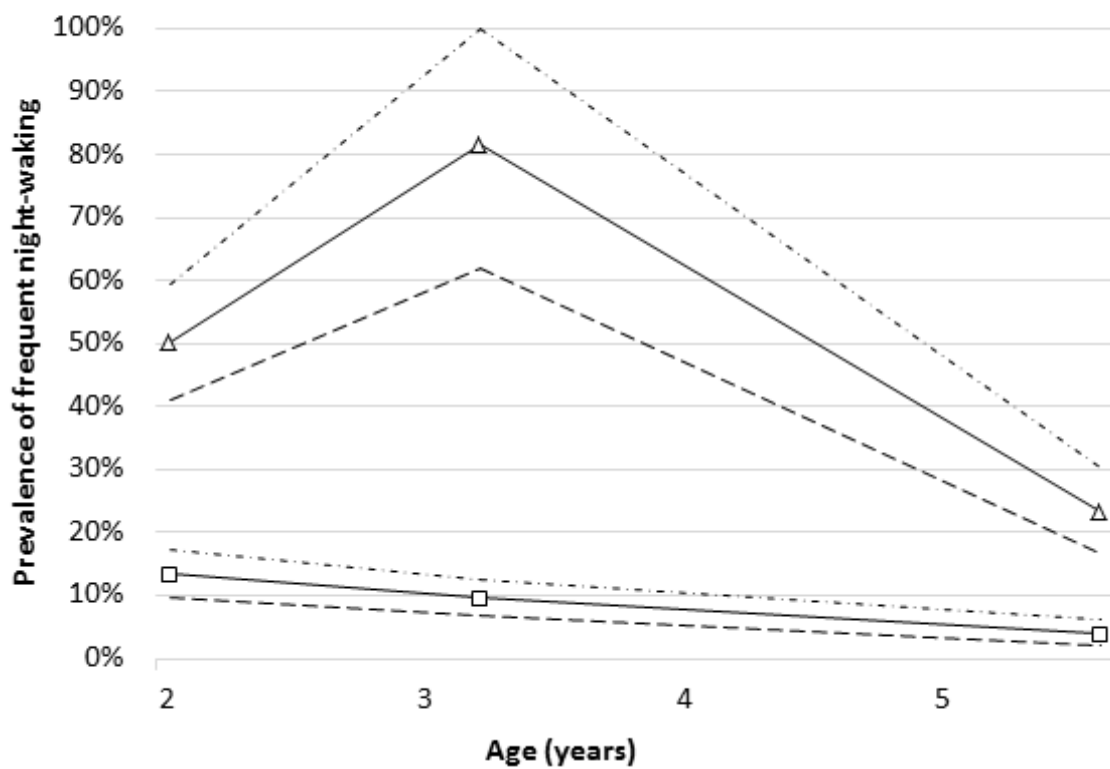

**a)** Frequent night-waking trajectories

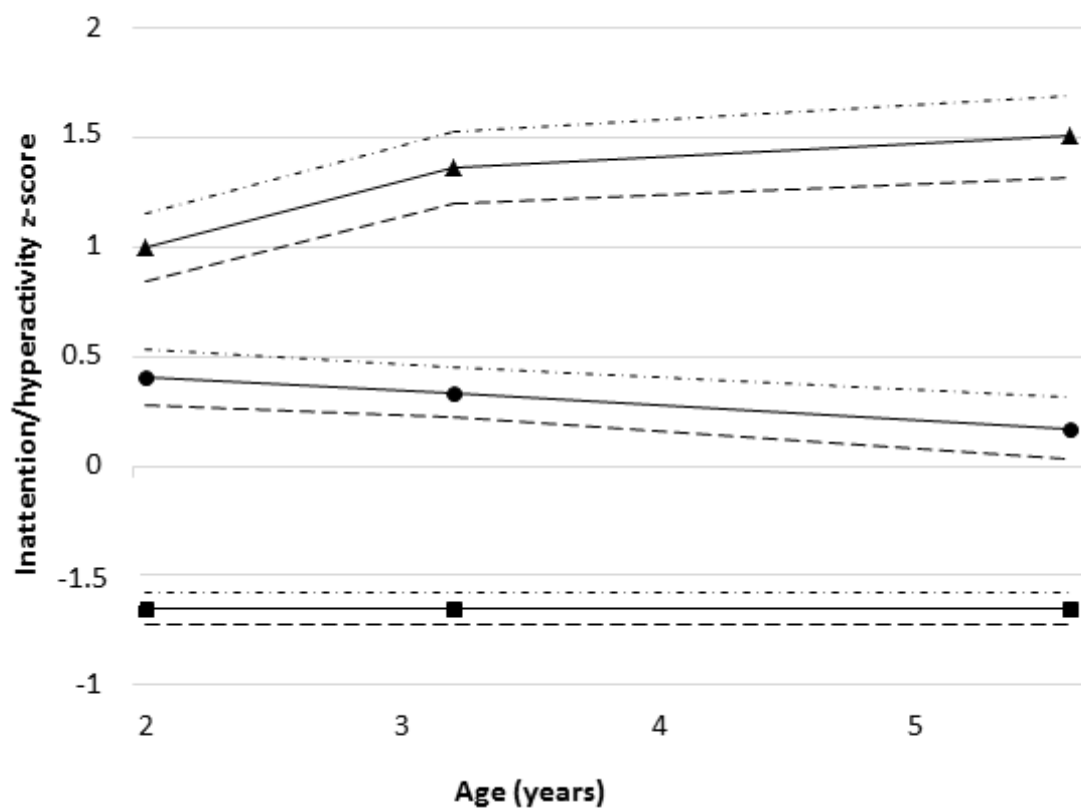

**b)** Inattention/hyperactivity z-scores trajectories

a) Distribution of children belonging to inattention/hyperactivity trajectories by night-waking trajectories

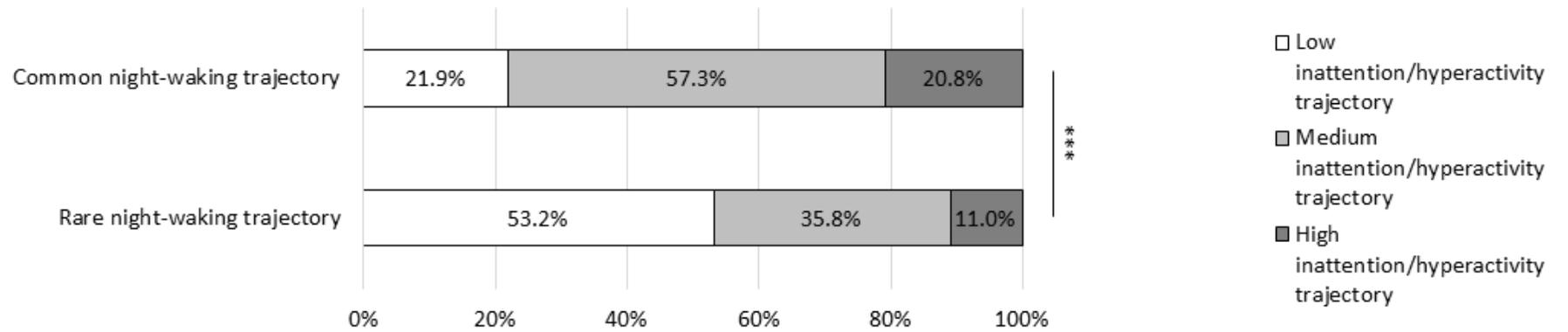

b) Distribution of children belonging to night-waking trajectories by inattention/hyperactivity trajectories

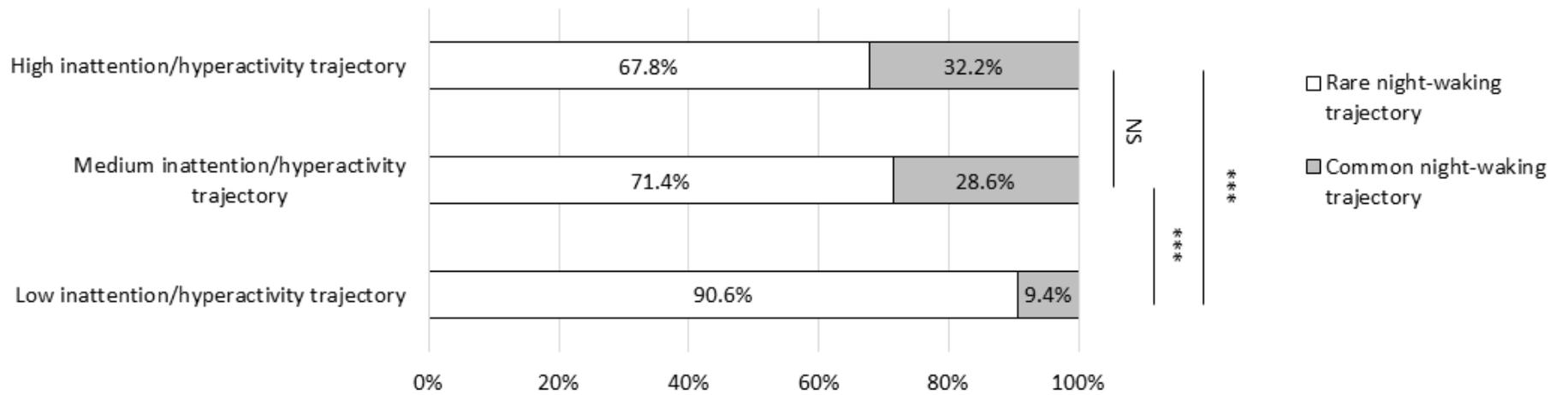

Figure 2. Prevalences of inattention/hyperactivity and night-waking trajectories

Supplementary data

Distribution of covariates according to joint trajectories (N=1342), reporting % (N) or mean (SD)

| %(N) / mean(SD) | Rare night-waking (N=1073) | | | Common night-waking (N=269) | | | P[b] |
|---|---|---|---|---|---|---|---|
| | Low I/H[a] N=571 | Medium I/H N=384 | High I/H N=118 | Low I/H N=59 | Medium I/H N=154 | High I/H N=56 | |
| **Parental characteristics** | | | | | | | |
| Household income | | | | | | | <0.001 |
|   <1500 €/month | 7.2% (41) | 12.2% (47) | 16.1% (19) | 6.8% (4) | 16.9% (26) | 26.8% (15) | |
|   1501-3000 €/month | 53.2% (304) | 61.5% (236) | 65.3% (77) | 72.9% (43) | 56.5% (87) | 62.5% (35) | |
|   >3000 €/month | 39.6% (226) | 26.3% (101) | 18.6% (22) | 20.3% (12) | 26.6% (41) | 10.7% (6) | |
| Education (years)[c] | 15.0 (2.3) | 14.2 (2.5) | 13.5 (2.3) | 15.1 (2.3) | 14.0 (2.5) | 13.4 (2.2) | <0.001 |
| History of childhood behavioral problem (yes) | 10.9% (62) | 16.7% (64) | 22.9% (27) | 11.9% (7) | 17.5% (27) | 25% (14) | <0.001 |
| Maternal depression (CES-D ≥23)[d] | 6.0% (34) | 7.6% (29) | 9.3% (11) | 5.1% (3) | 6.5% (10) | 7.1% (4) | 0.24 |
| Maternal age at delivery (year) | 30.6 (4.6) | 29.7 (4.9) | 29.1 (4.4) | 30.4 (4.3) | 30.0 (4.9) | 27.7 (4.5) | <0.001 |
| Smoking during pregnancy (yes) | 15.4% (88) | 23.4% (90) | 29.7% (35) | 17.0% (10) | 27.9% (43) | 25.0% (14) | <0.001 |
| **Child characteristics** | | | | | | | |
| Perinatal factors | | | | | | | |
|   Child gender (girl) | 53.4% (305) | 45.1% (173) | 28% (33) | 57.6% (34) | 45.5% (70) | 39.3% (22) | <0.001 |
|   First child (yes) | 44.1% (252) | 48.7% (187) | 50% (59) | 44.1% (26) | 42.9% (66) | 66.1% (37) | 0.043 |
|   Term at birth (weeks) | 39.3 (1.6) | 39.2 (1.8) | 39.4 (1.7) | 39.4 (1.3) | 39.1 (2) | 39.3 (1.7) | 0.34 |
| Lifestyle and sleep habits at age 2 | | | | | | | |
|   Collective care arrangement (yes) | 24.3% (139) | 22.4% (86) | 9.3% (11) | 28.8% (17) | 16.2% (25) | 12.5% (7) | <0.001 |
|   Television viewing (h/day) | 0.6 (0.6) | 0.8 (0.7) | 0.7 (0.8) | 0.8 (0.6) | 0.8 (0.7) | 0.9 (0.9) | <0.001 |
|   Inadvisable sleep habits | 8.4% (48) | 10.9% (42) | 7.6% (9) | 6.8% (4) | 13% (20) | 5.4% (3) | 0.48 |
|   Night-sleep duration (h/day) | 11.2 (0.8) | 11.1 (0.8) | 11.1 (0.7) | 10.9 (0.7) | 11.0 (0.8) | 10.9 (1.0) | 0.001 |

[a] I/H inattention/hyperactivity
[b] Global type 3 p-value obtained from unadjusted logistic regression, with the joint "rare night-waking and low inattention/hyperactivity trajectories" as reference
[c] Number of years of schooling starting from first year of primary school e.g. 12 years corresponds to having completed high school
[d] Center of epidemiologic studies depression scale, cutoff validated in a French population, measured during pregnancy